\documentclass[usenatbib]{basi}
\usepackage[T1]{fontenc}
\usepackage[british]{babel}
\usepackage[varg]{txfonts}
\usepackage{rotating}
\usepackage{dcolumn}
\begin{document}
\title[FRII RGs: jet dynamics and episodic behaviour]{FRII RGs: jet dynamics and episodic behaviour    }
%
\author[C.~Konar et~al.]%
       {C.~Konar$^1$\thanks{email: \texttt{chiranjib.konar@gmail.com}},
       M.~J.~Hardcastle$^{2}$ \\  
       $^1$Academia Sinica Institute of Astronomy and Astrophysics, NTU Campus, 
       Roosevelt Rd, Taipei-10617, Taiwan\\
       $^2$School of Physics, Astronomy and Mathematics, University of Hertfordshire, College Lane, Hatfield, UK}

\pubyear{2014}
\volume{00}
\pagerange{\pageref{firstpage}--\pageref{lastpage}}

\date{Received --- ; accepted ---}

\maketitle
\label{firstpage}

\begin{abstract}
Radio galaxies are episodic in nature. In our recent work, our study
of jet properties and their dynamics in different episodes of activity has
revealed various hitherto unknown aspects of the extragalactic jets.
We discover that the injection spectral indices are similar in the two
different episodes for most of the episodic radio galaxies in our
sample. We argue that in order to produce the similar injection
indices in two episodes (i) the jet power in different episodes have
to be similar and (ii) the Lorentz factor of the spine of the jet
should be > 10. We further argue from particle acceleration physics
that (iii) the jet fluid is made of electron-positron plasma and (iv)
the inner jets of double-double radio galaxies are capable of forming
hotspots even when propagating through the tenuous, nonthermal
electron-positron plasma of the outer cocoon without any thermal matter in
it. The episodic nature of the FR\,II radio galaxies we have studied
appears to be unrelated to AGN feedback on the ambient medium.
\end{abstract}


\section{Introduction}
\label{sec_intro}
Morphologically, Radio Galaxies (RGs) are of two types: Fanaroff-Riley
type I (FR\,I) and type II (FR\,II) \citep{Fanaroff.and.Riley-1974}.
FR\,II RGs are characterised by highly collimated jets and compact
hotspots at the outer ends of the lobes, whereas FR\,I jets are not
so collimated and have no hotspots. Here we confine ourselves to
FR\,II jets, in which the jet fluid flows relativistically up to the
jet termination point. RGs
are often episodic. When we find the observational signature of two
consecutive episodes of jet activity with a single host galaxy, we
call it a Double-Double Radio Galaxy (DDRG). In this paper, we
highlight (i) our study of FR\,II jet dynamics and (ii) how the study
of jet dynamics of the inner and outer doubles of DDRG FR\,IIs can throw
lights in solving many decades-old {\it open problems} related to RG
and jets.

\section{Observational results}
\label{sec_obsresult}
\begin{figure}
\centerline{
\includegraphics[height=4.5cm]{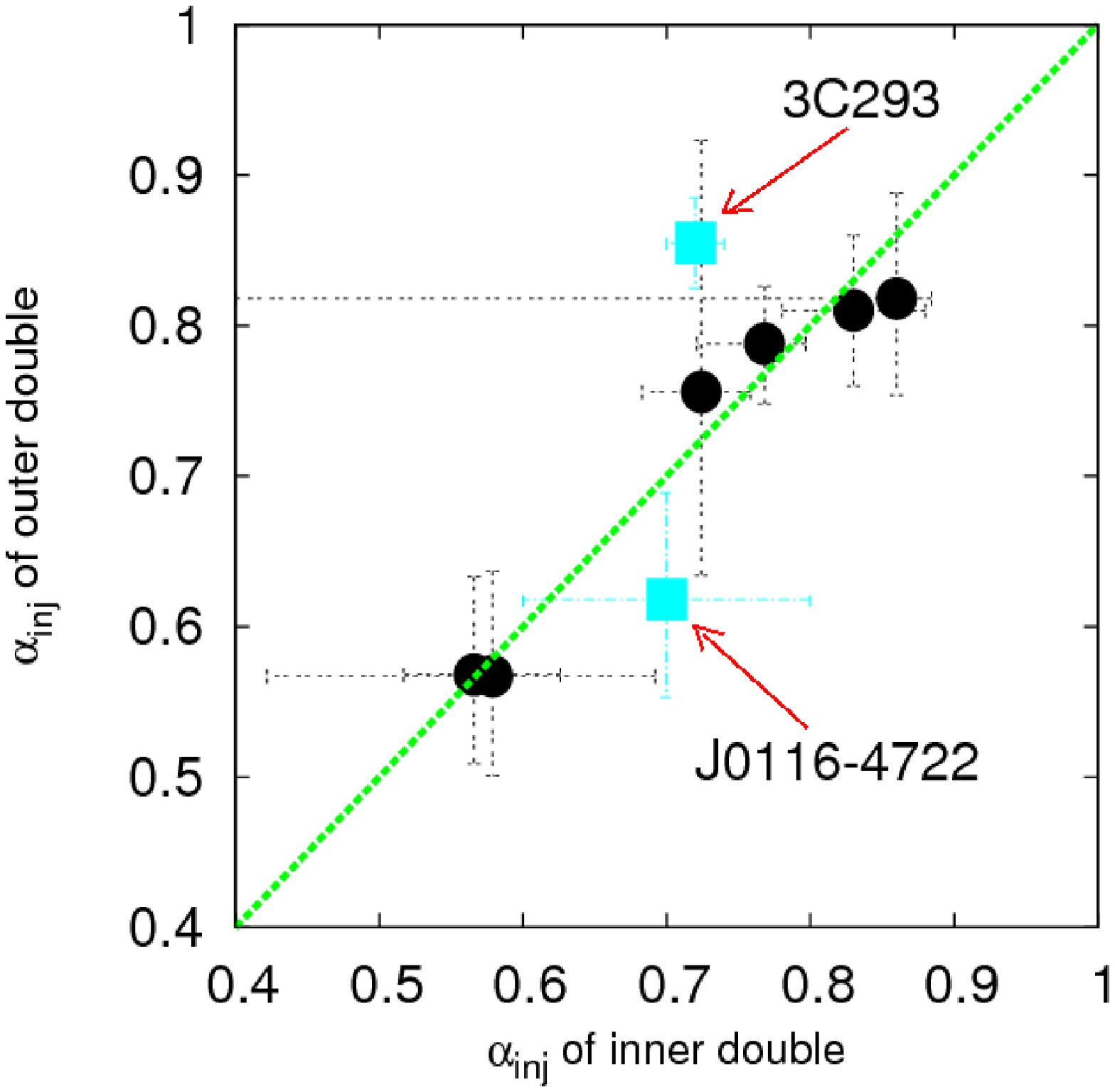}
\includegraphics[height=4.5cm,angle=0]{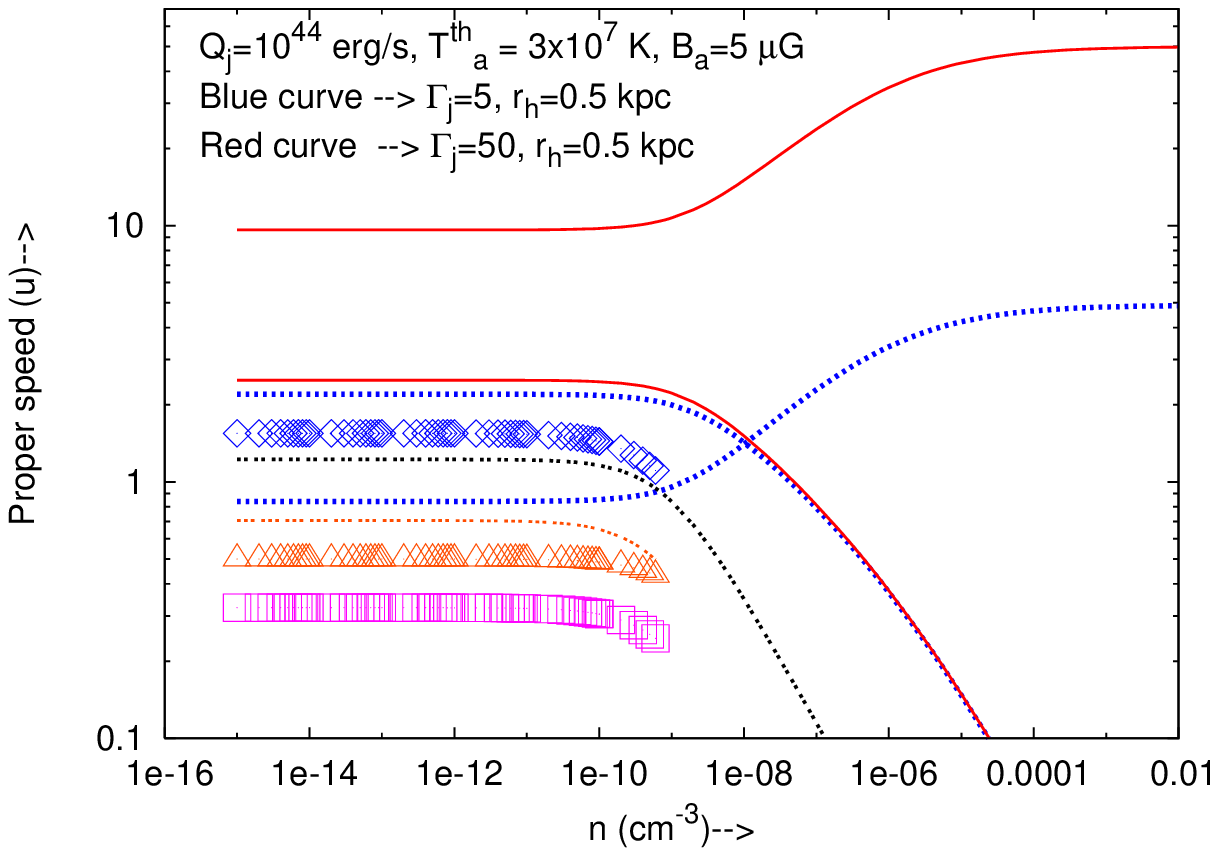}
}
\caption{Left panel: inner vs. outer injection index plot. Right
  panel: The variation of proper speed of the hotspot motion ($u_{\rm
    h}$) as measured in the host-galaxy frame and the upstream jet
  bulk motion as measured in the hotspot frame ($u_{\rm
    j,hs}=\Gamma_{\rm j,hs}\beta_{\rm j,hs}$) as a function of the
  number density ($n_{\rm p}$) of protons in the ambient medium. Some
  of the parameters used to draw the curves are annotated in the plot.
  The curves that fall and saturate as $n_{\rm p}$ decreases show the
  variation of $u_{\rm j,hs}$ with $n_{\rm p}$. The curves that rise
  (except the double-dotted black curve) as $n_{\rm p}$ decreases show
  the variation of $u_{\rm h}$. The black double-dotted curve shows
  the variation of the Alfv\'en speed. The orange triple-dotted curve is
  the sound speed in the absence of magnetic field (only in the
  ultrarelativistic regime). The other symbols indicate the variation
  of the magnetosonic waves mainly in the `pure nonthermal ambient
  medium', which are as follows: blue diamonds, the speed of the fast
  magnetosonic wave; orange triangles, the speed of the intermediate
  magnetosonic wave; and pink squares, the slow magnetosonic wave.
  These are published in \citet{Konar.and.Hardcastle-2013}.
\label{fig_1} }
\end{figure}

Our spectral ageing analysis of a small sample of 8 DDRGs (hereafter,
DDRG sample), compiled from \citet{Konar.etal-2006},
\citet{Konar.etal-2012}; \citet{Konar.etal-2013} and
\citet{Jamrozy.etal-2007}, revealed that (i) the injection spectral indices
(hereafter, injection indices) are similar in the two episodes of jet
activity (see Fig.~\ref{fig_1}) for our sample sources, (ii) the
duration of quiescent phase ranges from $10^5$ to $10^7$ yr and (iii)
there is a strong correlation between the injection index and jet
power for a small sample of FR\,II RGs (hereafter, FR\,II sample)
which includes the DDRG sample also. These samples are described by
\citet{Konar.and.Hardcastle-2013}. These results have important
implications in understanding the jet dynamics and cause of episodic
behaviour.

\section{Particle acceleration and composition of jet fluid}
\label{sec_jet-property}
An important questions is `What is a realistic model for the particle
acceleration phenomenon at the hotspots of FR\,II RGs?' We have shown
in our recent paper \citep{Konar.and.Hardcastle-2013} that the model
of \citet{Kirk.etal-2000} is fully consistent with our observational
results. This model deals with the particle acceleration at the
relativistic MagnetoHydroDynamic (MHD) shock in a fluid with
completely tangled magnetic field which is dynamically important.

It is believed that the particles are further accelerated at the Jet
Termination Shocks (JTS, i.e, hotspots). We argue that if the jet
fluid is an electron-proton plasma then by Drury's suggestion
\citep{Drury-1983} of `selectivity of injection', the protons, being
heavier, would be accelerated with more energy and in greater numbers
at the JTS before being injected into the lobes. So in the lobes, if
they are in equipartition at all, the magnetic field should be in
equipartition with the electrons and protons together. However, in
such a situation, the magnetic field would essentially be in
equipartition with the protons as they are energetically dominant,
which is contrary to observations \citep{Croston.etal-2005}.
Hence, we rule out significant numbers of protons in the jet fluid and
suggest that the jet fluid is made of pair plasma.

\section{Jet dynamics}
The momentum carried through the jet is transferred to the ambient
medium and the working surface (hence the hotspot) moves outwards due to
that. The hotspot motion is governed by ram pressure balance at
the working surface. For typical densities of the thermal ambient media
around RGs the hotspot motion is non-relativistic. However, the inner
hotspots of DDRGs can move relativistically (see
\citet{Konar.and.Hardcastle-2013} for observational evidence). So,
we need a relativistic ram pressure balance equation, given by
\begin{equation}
\beta_{\rm hs} = \frac{1}{1+  \sqrt{\frac{\beta_{\rm j}cA_{\rm h}}{Q_{\rm j}}w_{\rm a}} }\beta_{\rm j},
\label{eqn_mom.balance}
\end{equation}
where $\beta_{\rm hs}$ and $\beta_{\rm j}$ are the hotspot velocity
and jet bulk velocity in the HG frame, $A_{\rm h}$ is the area over
which the jet momentum flux is distributed, $c$ is the speed of light,
$Q_{\rm j}$ is the jet power and $w_{\rm a}$ is the relativistic
enthalpy density of the ambient medium surrounding the jet and lobes
(see \citet{Konar.and.Hardcastle-2013} for the derivation of this
equation). Our theoretical study of ram pressure balance shows that
for plausible values of the parameters occurring in
equation~(\ref{eqn_mom.balance}), (i) the motion of the inner hotspots
is supersonic relative to the first magnetosonic wave (hereafter,
first wave) of the cocoon matter and (ii) the bulk speed of the jet
fluid relative to the hotspot frame is faster than the first wave of
the jet fluid, even if the cocoon matter and the jet fluid are
ultra-relativistic non-thermal plasma. We conclude that
the inner jets can form JTSs as well as bow shocks under typical
observed conditions of DDRGs.

\section{Physical explanation of the results}
The particle acceleration model of \citet{Kirk.etal-2000} is the most
realistic one for the hotspots. So, a possible relation described by a
curve (in the plane of power-law index vs. upstream speed of the jet
fluid in JTS frame) of the kind shown in Fig.~4 of their paper, but
with an appropriate value of $\sigma_{\rm j}$ (i.e., $\sigma_{\rm j}$
at equipartition) with a very high Lorentz factor ($\Gamma_{\rm
  j}>10$) flow of jet fluid, combined with adiabatic loss and higher
synchrotron losses due to higher magnetic field at the hotspots of
high power sources, can explain the injection index$-$jet power
correlation as well as the similarity of injection index in two
different episodes. Also, we notice that for our FR\,II sample, there
is no obvious correlation between the injection spectral index and the
redshift of those sources. This enables us to conclude that jet
power$-$spectral index correlation is the primary one, and not the
redshift-spectral index correlation that has been discussed in earlier
work. The quiescent phase of our DDRG
sample is $10^5-10^7$ yr which is much smaller than the cooling time
of the typical ambient thermal medium of these DDRGs. This implies
that any feedback loop between the supermassive black hole and the thermal
ambient medium cannot be responsible for the episodic jet activity of
these sources.

\section{Conclusions}
The particle acceleration model of Kirk et al. (2000) is consistent
with the particle acceleration at the hotspots implied by
observations. This model can explain our observational results
presented in section~\ref{sec_obsresult}. Moreover, we have tackled
several decades-old problems which are as follows: 1) the jet fluid is
made of $e^-e^+$ plasma (we have given a completely new physical
argument), 2) the inner jets in a DDRG can form hotspots while
propagating through tenuous $e^-e^+$ plasma (which is contrary to
usual expectation), 3) Similar injection indices in two DDRG episodes
imply similar jet powers in the two episodes (we show this for the first
time), 4) we have concluded that $\Gamma_{\rm j} > 10$ to explain the
similarity of injection index in two episodes (this is surprising but
consistent with jet related X-ray as pointed out by
\citet{Hardcastle-2006}), 5) The $\alpha_{\rm inj}-Q_{\rm jet}$
correlation is the primary one and not the $\alpha_{\rm inj}-z$
correlation (we provide a plausible physical interpretation of this
correlation for the first time). These results are published in
\citet{Konar.and.Hardcastle-2013}.

\section*{Acknowledgements}
We acknowledge the organisers of the conference titled `Metre
Wavelength Sky' held in NCRA-TIFR, Pune, India.

\label{lastpage}
\end{document}